# Autonomous Systems – An Architectural Characterization

Joseph Sifakis

Univ. Grenoble Alpes, Verimag laboratory

## Abstract

The concept of autonomy is key to the IoT vision promising increasing integration of smart services and systems minimizing human intervention. This vision challenges our capability to build complex open trustworthy autonomous systems. We lack a rigorous common semantic framework for autonomous systems. It is remarkable that the debate about autonomous vehicles focuses almost exclusively on AI and learning techniques while it ignores many other equally important autonomous system design issues.

Autonomous systems involve agents and objects coordinated in some common environment so that their collective behavior meets a set of global goals. We propose a general computational model combining a system architecture model and an agent model. The architecture model allows expression of dynamic reconfigurable multi-mode coordination between components. The agent model consists of five interacting modules implementing each one a characteristic function: Perception, Reflection, Goal management, Planning and Self-adaptation. It determines a concept of autonomic complexity accounting for the specific difficulty to build autonomous systems.

We emphasize that the main characteristic of autonomous systems is their ability to handle knowledge and adaptively respond to environment changes. We advocate that autonomy should be associated with functionality and not with specific techniques. Machine learning is essential for autonomy although it can meet only a small portion of the needs implied by autonomous system design.
We conclude that autonomy is a kind of broad intelligence. Building trustworthy and optimal autonomous systems goes far beyond the AI challenge.

## 1. The concept of autonomy

The concept of autonomy is key to the IoT vision promising increasing integration of smart services and systems to achieve global goals such as optimal resource management and enhanced quality of life, with minimal human intervention.

This vision challenges our capability to build complex open trustworthy autonomous systems. In particular, we need an as much as possible, rigorous definition of autonomy. Is there a general reference model that could provide a basis for evaluating system autonomy? What are the technical solutions for enhancing a system's autonomy? For each enhancement, is it possible to estimate the implied technical difficulties and risks? These are very important questions for autonomous systems engineering.

Currently, the profusion of concepts and terms related to autonomy reflects the lack of a common semantic framework. It is remarkable that the technical discussion about autonomous vehicles focuses almost exclusively on AI and learning techniques while it ignores many other equally important autonomous system design issues.

What is the difference between a thermostat, an automatic train shuttle, a chess-playing robot, a soccer-playing robot and a robocar?

All bring solutions to the following general problem.

- A system consists of agents and objects sharing some common environment. It pursues a set of global goals to provide various services.
- Objects are physical dynamic systems without computation capability. Agents can partially observe and change their state. Objects can also undergo internal state changes.
- Agents have the ability to monitor the objects and act on their state, either alone or in some coordinated manner.
- The number of objects and agents can change dynamically depending on specific conditions.

*The problem is to determine the behavior of the system agents pursuing each one its own specific goals so that the collective behavior of the system including agents and objects meets given global goals.*

We propose a technical definition of autonomy based on a general computational model consisting of an agent architecture model and a system architecture model:

- The agent architecture model involves five modules, each one dealing with one fundamental aspect of autonomy: Perception, Reflection, Goal management, Planning and Self-adaptation. It specifies the coordination between these features in order to achieve autonomous behavior. It also implicitly defines some abstract partial order relation for comparing the autonomy level of agents pursuing identical goals.
- The system architecture model specifies coordination between system agents and their effect on the objects. We need such a model to explicate how an agent perceives its environment and elaborates its control strategy.

We progressively introduce the concept of autonomy through a comparison between five automated systems: a thermostat, an automatic train shuttle, a chess-playing robot, a soccer-playing robot and a robocar

## 1.1 Agent Environment

All the above systems automatically perform some mission characterized by their respective goals. They integrate agents continuously interacting with their environment through sensors and actuators. The sensors provide stimuli to the agent; the actuators receive commands from the agent and change accordingly the state of its environment.

 All agents receive inputs and produce outputs so that their I/O relation meets their specific goals. They are real-time controllers monitoring state changes of the controlled environment and producing adequate responses. Nonetheless, there are significant differences regarding the complexity and intricacy of their environments and their goals with associated decision process.

The environment of a thermostat is simply a room and a heating device. Stimuli are the temperature of the room and the state of the heater.
For the automatic shuttle, the environment includes the cars composing the shuttle with their equipment and passengers. Stimuli take the form of numeric information about the position and speed of the cars and the state of various equipment and peripherals.
For the chess robot, the environment is a chessboard with pawns and the adversary robot. Stimuli are the configuration of the pawns on the chessboard extracted from static images provided by the robot camera.
For the soccer robot, the environment consists of all other players, the ball, the goalposts and lines delimiting regions of the field. Stimuli are extracted in real time from dynamic images; they include the position and speed of players and ball.
Finally for the robocar, the environment is more involved as it includes vehicles and obstacles in its vicinity as well as traffic control and communication equipment. The perceived environment state is the configuration of other robocars and obstacles with their dynamic attributes and the state of traffic control and communication equipment. The environment state is built from data provided by different types of sensors adequately treated and interpreted.

## 1.2 Agent goals and plan generation

As explained, agents behave as controllers acting on their environment to achieve their specific goals. The agent environment can be modeled as a state machine with two types of actions: controllable actions triggered by the agent; and uncontrollable actions that are internal state changes of the environment. Without getting into technical details, given a set of goals and an agent environment model, there are methods (semi-algorithms) for the computation of plans. Figure 1 illustrates their principle for given environment and goals. In the considered example, the goals require that the generated plans avoid the state Bad and eventually reach the state Target. The plan generation method consists in finding a subgraph of the environment state graph that is closed with respect to uncontrollable actions (in red) and does not contain state Bad; furthermore, by adequately triggering controllable actions (in green) the state Target can be reached.
In general, environment models are infinite and the generated plans for given goals are infinite trees with alternating controllable and uncontrollable actions. When the environment model is finite, algorithms are used to compute a maximal controller including all the plans meeting the goals [1].

For infinite or complex environments, it is not possible to generate an explicit a controller. The existence of plans cannot be theoretically guaranteed. It depends on the type of goals and the controllability/observability relations.

In practice, for given goals, finite-horizon plans are computed on line from the agent's environment model. To cope with complexity, heuristics are used as well as precomputed plan skeletons. Furthermore, adequately choosing at design time the controllability/observability relation can significantly simplify on line plan generation. For instance, for simple safety goals e.g. avoiding harmful states, a finite horizon exploration from the current state may suffice.

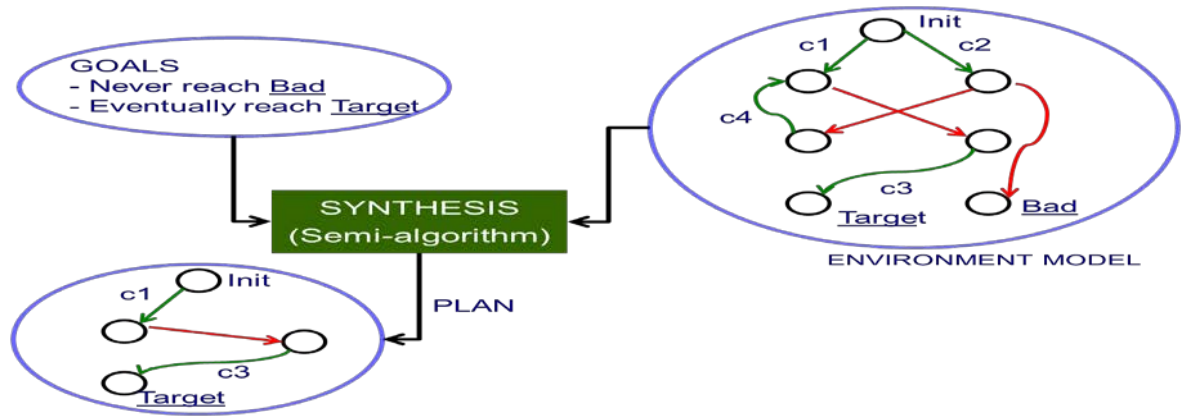

Figure 1: Plan generation from Goals and the Environment model

Going back to the considered examples, the thermostat has an explicit controller that is a simple two-mode automaton switching between *On* and *Off* modes when temperature reaches minimal and maximal values, respectively.

The shuttle has a more involved decision process. Usually an explicit controller ensures safety properties, while commands computed on line ensure adaptation to load variation and comfort optimization.

For the chess robot, there is no explicitly precomputed controller. Depending on the current configuration of the chessboard, the robotic agent chooses between a set of strategies optimizing criteria implied by the rules of the game. Each strategy corresponds to a sub-goal from a hierarchically structured set of goals. To accelerate plan generation, precomputed knowledge is often used e.g. patterns of plans and associated methods.

Similarly, for the soccer robot, plans are computed on line from the agent's environment model and its current configuration. Here a significant difference is the dynamic nature of the game as the controller is subject to hard real-time constraints. The game involves interaction between dynamically changing sets of agents (players). Although the game rules are well-defined, their dynamicity makes the outcome less predictable. The decision process generates plans from a dynamically changing environment model. It should adequately combine defense and attack strategies to win a game within 90 minutes. Knowledge is instrumental for plan generation; it consists in using precomputed patterns and learning techniques for parameter estimation.

For the robocar, the controller is even much more complex. In contrast to the previous examples, the environment involves a dynamically and unpredictably changing number of agents and objects in particular due to agent mobility. While for chess or soccer agents the gaming rules are static and well-understood, traffic rules are dynamic and hard to formalize [2]. Rigorous definition of a coherent set of individual goals for an ensemble of robocars is a non-trivial problem. Individual goals of robocars may be conflicting and a global consensus should be achieved in real-time taking into account multiple safety and optimality requirements.

## 1.3 Main aspects of autonomy

The discussed examples illustrate important differences when moving from simple automation to full autonomy. They also show technical obstacles to overcome in autonomous systems design. Autonomy is the capacity of an agent to achieve a set of coordinated goals by its own means (without human intervention) adapting to environment variations. It combines five complementary aspects:

- Perception e.g. resolution of stimuli, removing ambiguity/vagueness from complex input data and determining relevant information;
- Reflection e.g. building/updating a faithful environment run-time model;
- Goal management e.g. choosing among possible goals the most appropriate ones for a given configuration of the environment model;
- Planning to achieve chosen goals;
- Self-adaptation e.g. the ability to adjust behavior through learning and reasoning and to change dynamically the goal management and planning processes.

Note that the five aspects are orthogonal. The first two aspects deal with "understanding" the situation of the environment. The third and the forth aspect deal with autonomy of decision. Self-adaptation ensures adequacy of decisions with respect to the environment situation.

The above characterization, which we refine later, gives a clear insight about the very nature of the concept of autonomy. An autonomous agent uses at least one of the five functions. The thermostat is an automated agent that is not autonomous because its decision process is implemented by an explicit controller for a fixed set of goals. Furthermore, it has a fully observable/controllable environment providing stimuli that need no interpretation.

Automated agents are often integrated in complex processes where autonomy is ensured by human operators. For instance, PLCs ensure production automation while qualified staff performs supervision and overall coordination.

The level of autonomy of a system characterizes the relation between machine-empowered vs. human-assisted autonomy. Figure 2 illustrates this relation in a five-dimensional space. Improving autonomy for some aspect consists in replacing human intervention by autonomous steering. Full autonomy means that the function for each aspect is machine empowered.

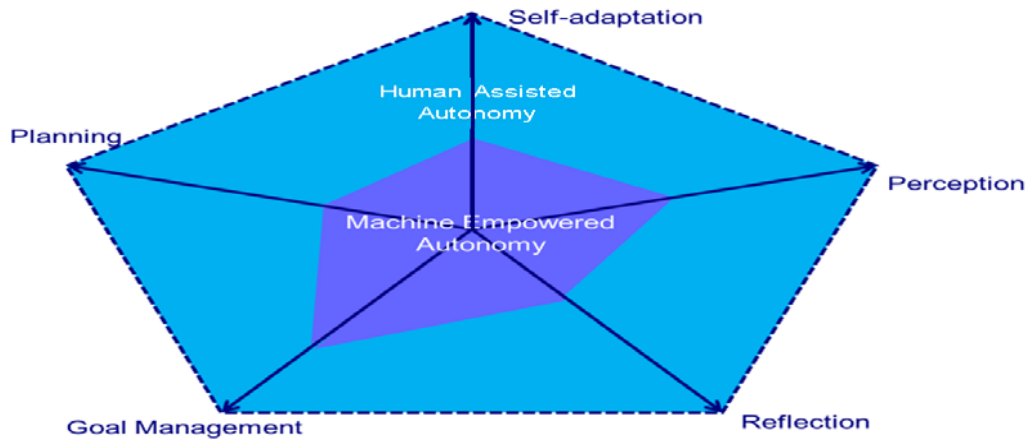

Figure 2: Human assisted vs. Machine empowered autonomy.

An illustration of this concept is provided by the five autonomy levels for cars defined by the SAE shown in Table 1. Level 5 corresponds to full autonomy while lower levels require increasing assistance of the driver.

| SAE AYTONOMY LEVELS |
|---|
| Level 0 | No automation |
| Level 1 | Driver assistance required ("hands on") |
| | The driver still needs to maintain full situational awareness and control of the vehicle e.g. cruise control. |
| Level 2 | Partial automation options available("hands off") |
| | Autopilot manages both speed and steering under certain conditions, e.g. highway driving. |
| Level 3 | Conditional Automation("eyes off") |
| | The car, rather than the driver, takes over actively monitoring the environment when the system is engaged. However, human drivers must be prepared to respond to a "request to intervene". |
| Level 4 | High automation ("mind off") |
| | Self driving is supported only in limited areas (geofenced) or under special circumstances, like traffic jams. |
| Level 5 | Full automation ("steering wheel optional") |
| | No human intervention is required e.g. a robotic taxi. |

Table 1: SAE autonomy levels (https://en.wikipedia.org/wiki/Self-driving_car)

## 2. A computational model for autonomous systems

### 2.1 A system architecture model

In order explain how an autonomous agent behaves, we need an adequate holistic model of its environment including other agents and objects. The model should in particular, propose concepts and principles accounting for the complex structure of the agent's environment and intricate coordination mechanisms.

We succinctly present an expressive architecture model developed with autonomy in mind. The model is inspired from the BIP coordination language. It has been studied and implemented in two formalisms, one declarative and another imperative [3,4,5].

As already explained, an autonomous system involves two kinds of components: agents and objects. The agents have computational capabilities. They can change the states of the objects and coordinate to enforce global system goals.

Components are instances of predefined types of agents and objects:

- An agent type is a computing system characterizing a mission or a service, e.g. Player, Arbiter, Sender, etc. Its semantics is a transition relation labeled with events and associated functions. Functions are triggered by the events that are atomic state changes involving other components, objects or agents.
- An object type is a dynamic system e.g. electromechanical system, whose state can change through interaction with other components. Note that some objects may be passive such as a pawn or a static obstacle.

We consider that a system model is a collection of architecture motifs, simply called <u>motifs</u>.

A motif is a "world" where live dynamically changing sets of agents and objects. It is equipped with a <u>map</u> represented by a graph specified by sets of nodes and edges. Nodes represent abstract coordinates in some reference space. The connectivity relation between the nodes of a map may admit a physical or a logical interpretation. For a lift or a shuttle, the map is a simple linear structure: the nodes are floors or stations, respectively. In the chess game, the map is an array representing the chessboard.

The position of an agent $a$ or of an object $o$ is given by a partial address function @: @($a$) and @($o$) is the node of the map where $a$ and $o$ are located, respectively.

For example, an address function can define the distribution of pawns over the chessboard. The function changes when pawns move; it is undefined for pawns not placed on the board. For the soccer game, the map is a three-dimensional array representing the field with some granularity grain. The only mobile object is the ball while all the agents are mobile.

Finally, for robocars we need several maps to model the system. Figure 3 depicts a model consisting of two motifs with their corresponding maps. A Road Chunk Map accounts for the spatial configuration of robocars and of relevant objects, typically obstacles. Other logical maps are necessary to specify coordination structure between robocars; for instance, to form platoons or to describe connectivity of communication infrastructure used by cars.

The dynamics of the system described by a motif is a transition relation between configurations. A configuration is the set of the states of its components as well as their corresponding addresses on the map. Configurations change when events occur as the result of agent coordination: by execution of interactions rules or of configuration rules.

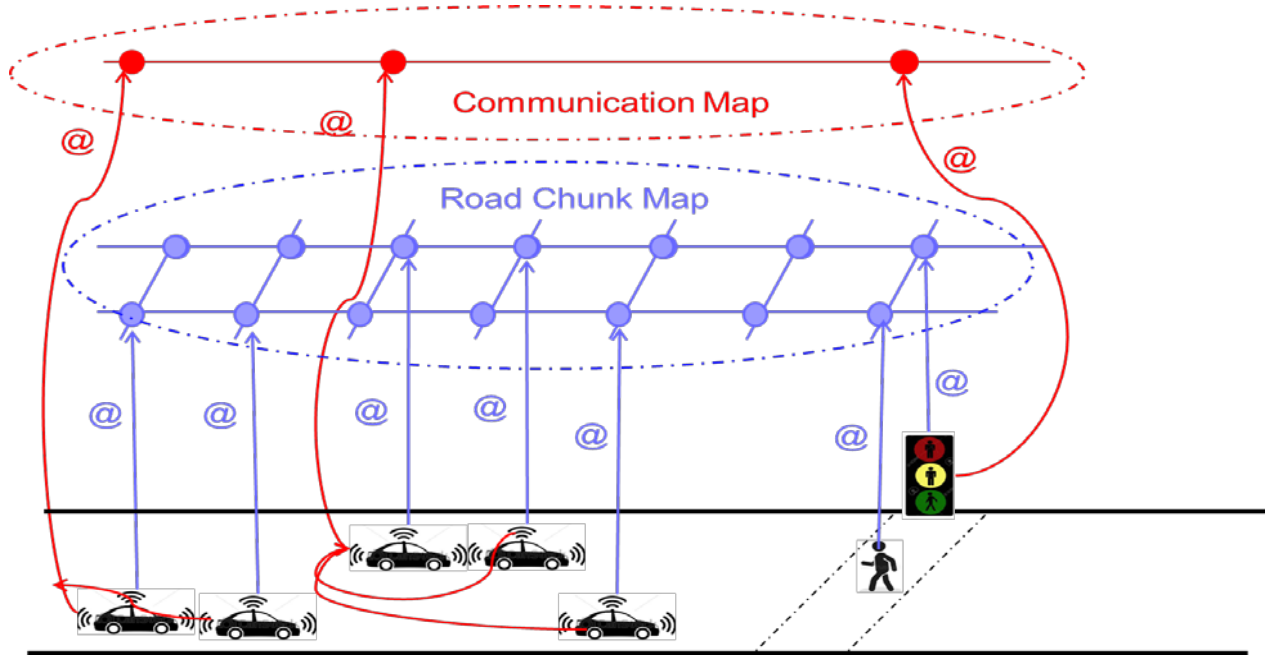

Figure 3: Modeling principle for robocars with two motifs

**Interaction rules**: An interaction is an atomic state change of a non-empty set of synchronizing agents that may also affect the state of objects. When the set is a singleton, the interaction is simply an action. We use rules in the form of guarded commands to describe interactions: the guard involves state variables of the synchronizing components and the command is a sequence of operations on their states. The rules are parametric which requires iteration over types of components. For example the rule

*for all a:vehicle, a':vehicle if [distance(@(a),@(a'))<l] then exchange(a.speed, a'.speed)*.

says that when two vehicles are close enough they exchange their speeds.

The model provides primitives encompassing strong or weak synchronization and interactions of arbitrary arity.

**Configuration rules**: Configuration rules allow the expression of three independent types of dynamism: component dynamism, component mobility, and map dynamism. They are guarded commands consisting of guards (conditions on state variables of components) and sequences of specific reconfiguration operations.

Typical operations are create/delete for components and add/remove for elements of maps. For instance, the operation *create(a:messenger,@(a)=n)* creates an agent named *a* of type *messenger* at address *n*. The operation *delete(o:pawn)* removes the pawn named *o*.

Agent mobility is modeled by rules modifying the address function of components. For example, the execution of the rule

*for all a:mobile if @(a)=n and @$^{-1}$(n+1)=empty then @(a):=n+1*

consists in moving forward agents of type *mobile* by one space of the map.

The proposed model is minimal and expressive. Each motif is a dynamic reconfigurable architecture, an ensemble of agents and objects governed by specific coordination rules.

Note that an agent may belong to more than one motif. Furthermore, components can migrate from one motif to another using reconfiguration commands.

For instance, the model of a soccer game involves at least two motifs.

The Attack motif ensures coordination rules that aim at getting inside the adversary's defense and finally score a goal. The Defense motif ensures coordination rules that aim at slowing down an offense to disrupt the pace and/or numerical advantage of an attack and finally get possession of the ball. Players can dynamically migrate from one motif to the other.

The model for an automated highway involves several motifs. All vehicles belonging to a Road Chunk motif are subject to general traffic coordination rules. A Platoon motif groups and coordinates an ensemble of vehicles cruising at the same speed and closely following a leader vehicle. An Overtake motif involves an overtaking vehicle and vehicles moving in the same direction in its vicinity. Finally, a Communication motif groups vehicles sharing a common communication infrastructure.

## 2.2 A computational model for agents

We present the agent computational model that puts emphasis on architectural aspects following the same line as [7]. It consists of four main modules and a Repository as depicted in Figure 4.

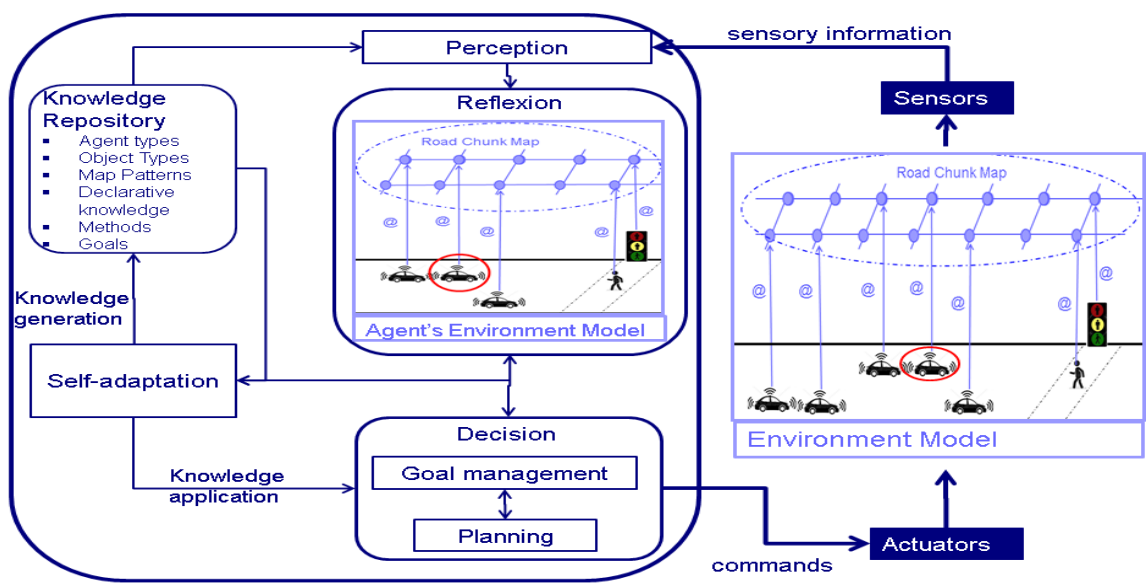

Figure 4: The general architecture of the computational model for agents

### 2.2.1 The Knowledge Repository

The Knowledge Repository contains different kinds of knowledge used by the other modules for 1) the interpretation of sensory information; 2) building the environment model of the agent; 3) goal management and subsequent goal planning.

Some of the Repository knowledge is developed at design time and some is produced and stored at run time.

Design time knowledge specifies basic components of the agent's environment, their main observability/controllability features as well as key properties and methods related to system goals. It includes in particular:

- A list of all the relevant types of agents and objects and their corresponding behavioral specification with the admitted coordination patterns e.g. interaction types and reconfiguration commands;
- A list of predefined maps and coordination patterns used to build the agent's environment model;
- A list of the goals pursued by the agent as sets of properties of two types: 1) critical properties requiring that some condition is never violated; 2) best-effort properties dealing with resource optimization e.g. finding tradeoffs between performance and resource utilization.
- A list of methods used to enrich the knowledge about the environment model and so to produce additional knowledge at run time e.g. monitoring and learning techniques.

Run time knowledge is generated on line from monitors, learning and analysis techniques. It includes in particular:

- Properties of the agent model that may be generated by application of analysis techniques or inferred by application of reasoning techniques;
- Knowledge produced by monitors of the agent's behavior e.g. detecting failures or intrusion;
- Knowledge produced by application of learning techniques, in particular to remove ambiguity about the environment configuration or to estimate parameters characterizing the dynamic behavior of the environment e.g. worst-case and average execution times.

This presentation leaves open important questions about the nature of knowledge and the different forms it can take [2]. We discuss below some issues relevant for agent design.
We consider that knowledge is "truthful" information that is used in some specific context to understand/predict a situation or to solve a problem. Truthfulness cannot always be asserted in a rigorous manner. Mathematical knowledge has definitely the highest degree of truthfulness, e.g. knowledge extracted from programs using analysis tools. At the other extreme, empirical knowledge although not theoretically substantiated, proves to be very useful in practice. The most widely used knowledge is empirical e.g. common sense knowledge, but also knowledge from machine learning.

Knowledge may be declarative or procedural, regarding the form it can take. Declarative knowledge is a relation (property) involving entities of a domain. In the Repository, can be stored: 1) logic formulas inferred from a set of axioms; 2) valid system properties extracted from a system model e.g. system invariants; 3) architecture patterns enforcing given properties.
Procedural knowledge takes the form of an executable description such as algorithms, behavioral description of components and various analysis techniques.
The Knowledge Repository contains all these types knowledge. Utilizing them effectively is essential for ensuring agent's self-adaptation and autonomy.

### *2.2.2 The Perception module*

The Perception module extracts relevant information from the various stimuli provided by sensors. For this purpose, it makes use of learning techniques or of analysis and recognition processes. The extracted information is linked to knowledge of the Repository. It concerns

- the type and possibly the identity of each sensed agent or object;
- the state of the so identified components;
- the type of the external environment characterized as a set of motifs with maps and associated coordination features.

For instance, the Perception module of a soccer agent provides, for each identified component of the environment, its position and speed in the field map. The Perception module of a robocar provides the types of the components in the vicinity with their associated attributes. Some attributes connect the components to motifs and their corresponding maps.

### *2.2.3 The Reflection module*

The Reflection module uses information provided by the Perception module in order to build/update a model of the agent's environment. For some agents, the environment model - number of components, map, coordination rules - does not change over lifetime e.g. chessboard robot, soccer robot. Thus, sensory information determines mainly the state of components e.g. their position in the maps and interactions.

Agents with dynamically changing environment e.g. robocars, are initially equipped with some environment model that is dynamically updated e.g. by creating/deleting motifs. For this to be feasible, the stimuli should provide information about architectural changes of the environment. Furthermore, the detected changes should correspond to patterns stored in the Knowledge Repository.

Reflection module extensively uses design-time knowledge of the Repository to build a complete behavioral model of its perceived environment. Nonetheless, to preserve faithfulness and freshness of the model, stimuli interpretation should be precise enough and performed within acceptable delay.

Performance of this module is critical for mobile agents subject to real-time constraints. How fast the agent's environment model can track changes of the real environment? Additionally, for distributed multi-agents systems, there is an inherent uncertainty about the global system states and thus a risk of discrepancy between environment models of different agents [7].

Note that each agent builds a partial model of the system environment reflecting its knowledge about its "neighborhood" that can be observed. In a distributed system, there is no global model of the system environment.

### *2.2.4 The Decision module*

The Decision module is decomposed into two cooperating submodules: a Goal Manager handling the actual agent's goals and a Planner generating plans that implement particular goals.

The module manages a set of goals both critical and best effort. It assigns higher priority to critical goals according to their importance.

Often goal management boils down to solving an optimization problem. It consists in translating goals into utility function policies: a goal is characterized as the desired set of feasible states for which the objective function is optimized subject to a set of constraints [9].

For a selected goal, the Planner computes from the environment model a corresponding plan. To cope with the exploding complexity of the planning process, various heuristics and precomputed patterns from the Knowledge Repository may be used.

The generated plans involve commands for interaction with other agents or reconfiguration of their environment as explained in the system architecture model. The allowed coordination patterns with other components of the environment are specified in their definition stored in the Knowledge Repository. Note that interactions may involve exchange of knowledge between interacting agents e.g. changing methods or goals.

### 2.2.5 The Self-adaption module

The Self-adaptation module supervises and coordinates all the other modules. It continuously reassesses the coherency of the exchanged information, creates new knowledge and provides directives to the Goal manager.

The module applies existing knowledge or generates new knowledge by combining reasoning and run-time analysis techniques to detect significant changes in the environment that require responsive adaptation. For instance, it applies monitoring or analysis techniques to the environment model to detect critical situations; it also can use learning techniques to estimate parameters or detect abnormal situations.

The adaptation directives to the Goal Manager concern:

1) Change of parameters affecting the choice of the managed goals, especially estimates of dynamic characteristics of the environment components;

2) Change of the set of the managed goals (adding or removing a goal), in response to some exceptional event in the environment or to an explicit requirement through interaction with another agent.

## 3. Autonomous system design complexity issues

An interesting technical question is how to adequately choose the autonomy level for risk-benefit optimization in system design. Four main factors determine this choice.

The first is the required degree of trustworthiness. For critical complex systems, semi-autonomy seems to be the realistic choice under the current state of the art e.g. ADAS cars.

The other factors are three independent types of complexity discussed below: autonomic complexity, design complexity and implementation complexity.

## 3.1 Autonomic complexity

We need a concept of complexity accounting for the specific difficulty to build autonomous systems. The following factors related to the fundamental aspects of autonomy, capture autonomic complexity.

1. <u>Complexity of perception</u> characterizes the difficulty to interpret stimuli provided by the environment and to timely generate corresponding inputs for the agent environment model. It has various sources such as stimuli ambiguity (admitting different interpretations) or vagueness (fuzzy or noisy stimuli). Additionally, complexity is aggravated with the volume of stimuli data to be analyzed in order to extract relevant input information.
2. <u>Lack of observability/controllability</u> which implies partial knowledge of the agent's environment and consequently limitations for building a faithful run time model by the Reflection module. This affects the ability to build plans and act on the environment.
3. <u>Complexity of goal management</u> which is the complexity of the process of choosing amongst a set of goals a maximal subset of compatible goals characterizing a strategy for which a consistent plan is generated. The selection process may involve both qualitative criteria such as priorities and quantitative criteria such as optimization of physical quantities.
4. <u>Complexity of planning</u> which directly depends on the type of goals and the complexity of the agent's environment model. As explained goals may be as simple as non-violation of a constraint and more complicated such as reachability of a condition or achieving optimality over a given time period.
5. <u>Complexity of adaptation</u> which is directly related to uncertainty about the agent's environment. Sources of uncertainty are multiple, including time-varying load, dynamic change due to mobility, bursty events, and most critical events such as failures and attacks. The Self-adaptation module generates objectives to cope with such situations involving imperfect knowledge and lack of predictability [2,5]. This can be achieved to some extent, using knowledge, e.g.[8].
Note that reduced observability is a source of uncertainty. Nonetheless, uncertainty is not completely resolved by simply enhancing observability [2].

Note that for agents not directly interacting with a physical environment, autonomy simply means to cope with the complexity of goals and some uncertainty e.g. an encoder adapting to varying load to avoid frame skipping. For a chess robot, only complexity of goals and planning are relevant; its environment is fully observable/controllable without uncertainty and the stimuli are non-ambiguous. For robocars, all types of complexity are relevant.

## 3.2 Design complexity and its relationship to autonomy

System design complexity characterizes the difficulty to build a system out of components – autonomous or not. It is conceptualized in a two-dimensional space [2].
One dimension represents reactive complexity [10] of the agents constituting a system.
The other dimension represents the complexity of the architectures used to coordinate the agents.

Although design complexity is independent from autonomic complexity, it is interesting to understand how the demand for autonomy affects system design choices.

### *3.2.1 Reactive complexity*

Reactive complexity characterizes the intricacy of the interaction between an agent and its environment. It is independent from space complexity or time complexity measuring the quantity of computational resources needed by an agent.

We discuss below a classification of agents according to their reactive complexity (Figure 5).

- The simplest agents are transformational agents where the relation of the input to the output is sufficient to characterize their behavior. Computation is performed in batch mode without reference to any operating environment. Such agents are often software systems oblivious to real-time constraints, with simple well-defined environments. Adaptation consists in using precomputed knowledge to cope with inherent complexity of decision problems, e.g. intelligent resource orchestration in data centers, intelligent personal assistant, game playing agent.

- Streaming agents compute functions on streams of data. For a given input stream of values, they compute a corresponding output stream. The output value at some time *t* depends on the history of input values received by *t*. The goals for streamers deal with functional correctness and specific time-dependent properties such as latency. Data-flow systems are usually composed of streamers. Adaptation is essential to cope with load unpredictability and meet latency constraints, see for example [11]

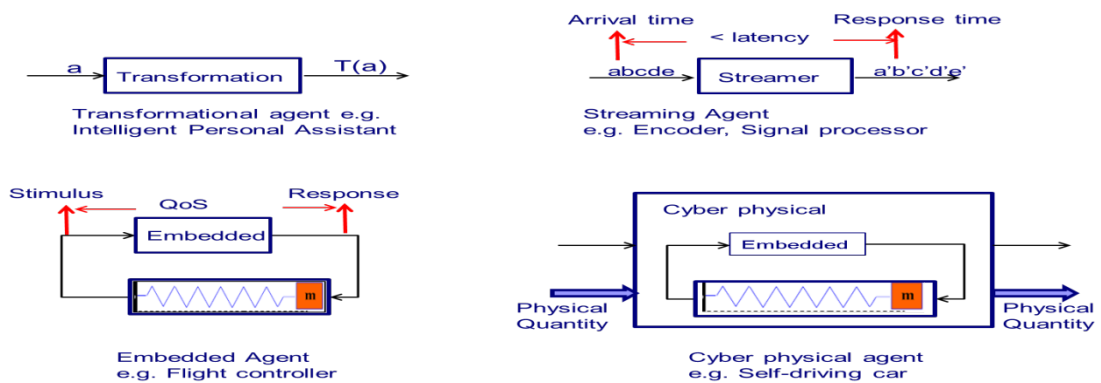

Figure 5: Classification of agents according to their behavioral complexity

- Embedded agents continuously interact with a physical environment to ensure global properties. They are mixed HW/SW systems where real-time behavior and dynamic properties are essential for correctness. Autonomous behavior is required when their mission involves high-level goals and complex environments, in particular to adaptively manage computational resources and meet critical goals. Embedded agents are integrated in industrial systems, transport systems and all kinds of devices.
  Note that the model of embedded agents should account for the behavior of their internal environment including computational resources (see discussion below).

- A cyber physical agent is an embedded agent integrating in its internal environment objects that are exclusively under its control. Its behavior involves both discrete and continuous variables representing the state of the integrated objects.
  The environment model of such an agent should be refined to distinguish between internal and external environment as shown in Figure 6. The Perception module gets sensory information from both the external and the internal environment model. The Reflection module builds/updates the two models corresponding to the two environments. The decision process is applied to the product of the environment models to generate plans with commands acting on both environments.

Cyber physical systems seek a tight integration between computers and their physical environment. They are essential for building complex autonomous systems e.g. self-driving cars.

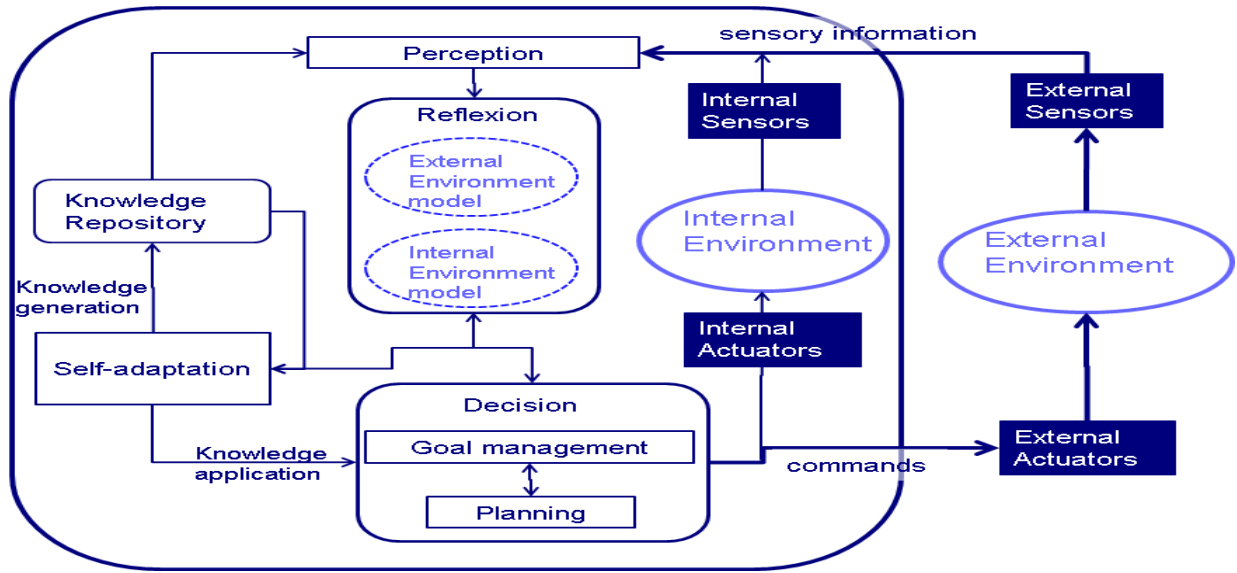

Figure 6: Computational model for cyber physical agent

### 3.2.2 Architecture complexity

The proposed model in 2.1 provides a basis for classifying system architectures according to their degree of dynamism, from static to self-organizing architectures as shown in Figure 7.

We enumerate some representative cases below for increasing complexity of coordination.

a) <u>Static architectures</u> involve a given number of agents and objects, with fixed coordinates e.g. a smart building with fixed microcontrollers and electromechanical equipment.

b) <u>Parametric architectures</u> can have arbitrary initially known numbers of "pluggable" components for fixed coordination patterns e.g. token ring architecture, an array computer.

c) <u>Dynamic architectures</u> are parametric architectures with dynamic creation/deletion of agents or objects, e.g. array architecture for the Game of Life, client-server architecture.

d) <u>Mobile architectures</u> are dynamic architectures where also the coordinates of objects and agents can change dynamically, e.g. swarm robotic system. Additionally, they may involve dynamic change of maps when mobile agents explore a space and progressively build a model of their environment.

e) <u>Self-organizing architectures</u> are mobile architectures with many dynamically changing motifs e.g. for robocars, soccer playing robots. Self-organization is the ability to adapt coordination rules to changing system dynamics.

| System Architecture Model | Motifs | Single motif | Many motifs | Dynamic motifs |
|---|---|---|---|---|
| | Interactions | No agent interaction (actions on objects) | Static interactions | Parametric interactions |
| | Reconfiguration — Dynamic agents | | NO | YES |
| | Reconfiguration — Dynamic objects | | NO | YES |
| | Reconfiguration — Mobility (dynamic @) | | NO | YES |
| | Reconfiguration — Dynamic map | | NO | YES |
| System Goals | | Single fixed goal | Many structured goals | Many conflicting goals | Dynamic goals |
| Implementation | Agent | Software | Hardware/Software | Cyber physical |
| | Architecture | Centralized | Decentralized | Distributed |

Figure 7: Variation of system complexity with respect to architecture, goals and implementation

We have shown that all these types of architectures can be formalized as operators taking as arguments arbitrary numbers of instances of agent and objects types [3,4,5]. We badly need theory for studying their properties in a compositional manner. Knowing the properties of the types of objects and agents involved, is it possible to infer global system properties? A more ambitious avenue is to develop theory for correctness by construction [6]: how to combine basic architecture patterns with well-established properties in order to build complex architectures that preserve the properties. These are largely open hard problems that urgently need exploration.

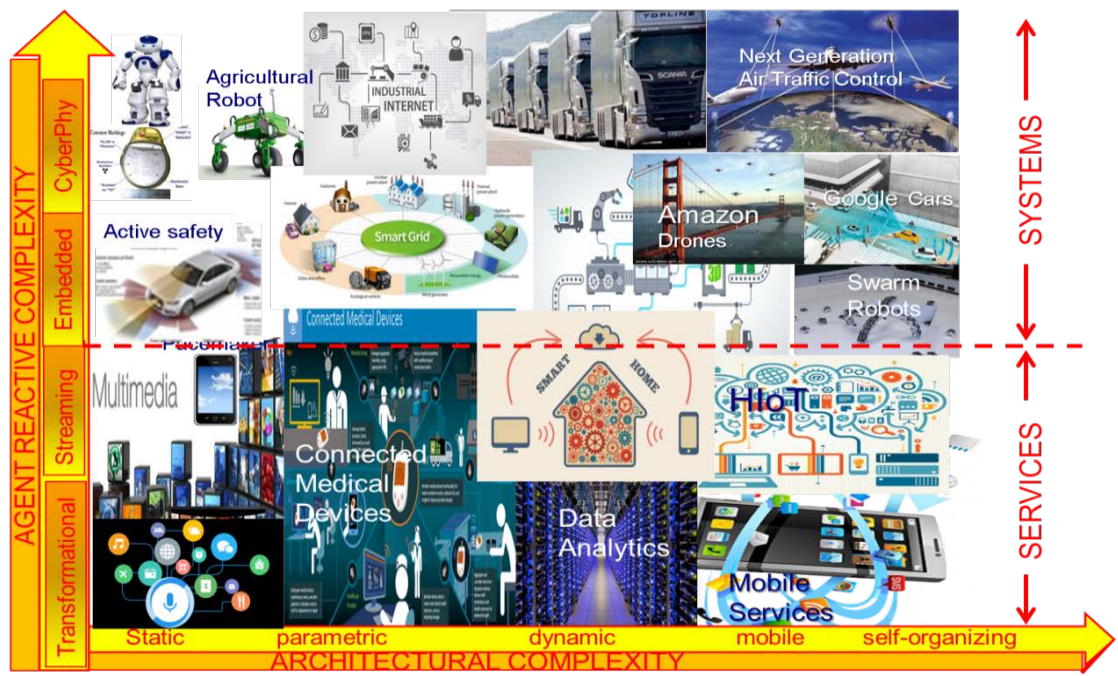

Figure 8: Design complexity

Figure 8 illustrates design complexity for different types of systems depending on the reactive complexity of their agents and their architectural complexity. Note the separation between services and systems. Services use streamers and transformational agents. IoT systems with advanced autonomy features, require mobile or self-organizing architectures and integrate embedded or cyber physical agents. Self-organization is important for such systems with many conflicting goals. Nonetheless, contrary to common opinion, self-organization is not an intrinsic property of autonomous systems. An ordinary distributed system involving agents with explicit controllers communicating by exchange of non-ambiguous messages is self-organizing if it has multiple coordination modes. Similar arguments are applicable for other "self"-prefixed properties commonly considered as characteristic properties of autonomous systems.

### 3.3 Implementation complexity

Implementation is the process that leads to the realization of the designed system model. The latter can admit different implementations depending on the available computational resources and their organization. In rigorous approaches, the outcome of the implementation process is another model accounting for the physical distribution of agents and features of infrastructure implementing the model coordination mechanisms [12].

We discuss below main choices for the implementation architecture depending on how/where the decisions are made and how/where the information is shared between the coordinating agents. We distinguish three main types of implementation architecture.

1. Centralized architecture where the agents are not geographically distributed. They coordinate through a shared memory that stores the data of a common Knowledge Repository as well as the data representing the state of a common environment model. In other words, each agent directly modifies/reads a shared data structure representing the motifs with their maps and the associated addressing functions. Such an implementation presents the advantage of the overall coherency of decision and coordination. Nonetheless, access conflicts may affect performance. A typical example is a blackboard architecture equipped with a common knowledge base, iteratively updated by agents starting with a problem specification and ending with a solution.

2. Decentralized architecture where agents are geographically distributed and there is no central storage. Every agent makes decisions based on local knowledge and the resulting system behavior is the aggregate response. Nonetheless, agents can coordinate through local memory depending on the topology of the environment maps. A typical example are stigmergic systems where mobile independent agents e.g. ants, robots, use their common environment to for coordination purposes [13]

3. Distributed architecture where there are no shared data storages. Each agent handles its own data and makes decisions according to its own goals. Coordination between agents is exclusively through asynchronous message passing. A key issue for such systems is coherency of coordination between components to achieve global goals. These are an emerging property of the collective behavior of the agents.

Distributed autonomous agent systems are today a vast and active research field because of multiple applications in various domains from blockchain protocols to complex autonomous transportation systems.

# 4. Trustworthy autonomous systems – From correctness at design time to autonomic correctness

Systems Engineering comes to a turning point moving from small-size centralized non-evolvable automated systems with predictable environments, to large distributed evolvable autonomous systems with non-predicable dynamically changing environments.

Is it possible to build trustworthy autonomous systems? As autonomous systems are often critical, this is the object of a considerable and sometimes heated debate [15]. As explained in [2], the trend for autonomous systems renders obsolete current critical systems engineering techniques and standards, such as ISO26262 and DO178B, that require conclusive trustworthiness evidence based on some rigorous design methodology.

It is remarkable that currently cars with autonomy features are self-certified by their manufacturers, contrary to most industrial products that are certified by independent authorities. Furthermore, some carmakers consider that successfully passing an extremely large number of test cases is a sufficient evidence of trustworthiness.

Trustworthiness is a transversal design issue. It is not limited to purely functional correctness. A system is deemed trustworthy if it behaves as expected despite design errors, hardware failures and any kind of harmful interaction with its human and physical environment, including misuse, attacks, disturbances and any kind of unpredictable events [6].

We briefly discuss how the rigorous model-based approach for guaranteeing trustworthiness can be in principle, extended to autonomous systems and the implied technical difficulties.

Currently, model-based approaches for achieving trustworthiness involve two steps.

The first step aims at providing underlined guarantees that some abstract system model representing the system's nominal behavior satisfies critical system goals. The nominal behavior model usually assumes that system environment is fully reliable and to some extent predictable. The second step deals with possible violations of these assumptions for a given implementation.

Building autonomous system models accounting for nominal behavior requires strong expertise on both modeling and algorithmic aspects. Algorithms describe how individual goals of agents contribute to achieving global system goals. Their design is a non-trivial problem because they are distributed or decentralized. Furthermore, they pursue jointly critical and optimization goals for dynamically changing environments. They allow the management of critical resources (space, time, memory, energy) by optimizing performance and additionally respecting smoothness conditions. Typical examples are collision avoidance algorithms for vehicles (cars, aircraft) that manage the available space respecting requirements on speed and avoiding collision with obstacles. Other examples are mixed criticality systems involving critical and non-critical features.

Modeling deals with agent nominal behavior description and coordination. Agent nominal behavior assumes that both the sensors and the Perception function are flawless and that sensory information is correctly interpreted into predefined concepts. It focuses on Reflection and Decision and in particular on their dynamic aspects.

Following our approach, the coordination is described as the composition of motifs each one corresponding to a system mode and solving a specific coordination problem. Model correctness can be inferred in principle, by proving that the motifs are correct with respect to their coordination goals and that they are composable [6].

Providing guarantees for complex autonomous systems faces several limitations [2]. One is the decomposition and formalization of high-level goals in terms of concrete requirements verifiable on the system behavioral model. A second limitation concerns our ability to build faithful system models, especially when they involve cyber physical components. The third limitation is that machine-learning techniques do not lend themselves to behavioral modeling and should be treated as "black boxes".

The second step aims at ensuring trustworthiness for a given implementation taking into account deviations from nominal behavior e.g. possible harmful events such as failures and security threats. It starts from the characterization of trustworthy states for nominal behavior provided by the first step (Figure 9). It involves a more or less exhaustive analysis to identify all kind of harmful events and their possible effect. Then, for each harmful event, specific techniques are used to ensure resilience e.g. typically, redundancy-based techniques. This practically means that the occurrence of a single harmful event does not (immediately) compromise system trustworthiness. It leads to some non-fatal state from which using DIR (Detection, Isolation, Recovery) mechanisms it is possible to bring the system back to a trustworthy state [14].

This approach has been successfully applied to small, centralized critical systems. It is costly and leads to overprovisioned systems [6] as it consists in estimating independently, for each type of harmful event and associated DIR mechanism, worst-case situations and statically reserving the needed resources to cope with them. Its application to autonomous systems is even more difficult as the characterization of the effect of harmful events depends on complex environmental conditions. Such a characterization cannot be enumerative and exhaustive; it should be symbolic and conservative, the result of a global model-based analysis.

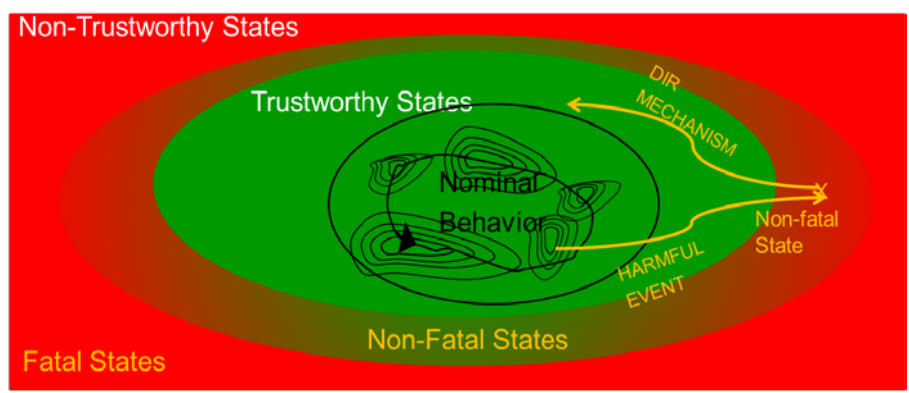

Figure 9: Recovery from non-fatal states

| | | | |
|---|---|---|---|
| 1 | Vehicle Failure | 19 | Vehicle(s) Drifting – Same Direction |
| 2 | Control Loss With Prior Vehicle Action | 20 | Vehicle(s) Making a Maneuver – Opposite Direction |
| 3 | Control Loss Without Prior Vehicle Action | 23 | Lead Vehicle Accelerating |
| 4 | Running Red Light | 24 | Lead Vehicle Moving at Lower Constant Speed |
| 5 | Running Stop Sign | 25 | Lead Vehicle Decelerating |
| 6 | Road Edge Departure With Prior Vehicle Maneuver | 26 | Lead Vehicle Stopped |
| 7 | Road Edge Departure Without Prior Vehicle Maneuver | 27 | Left Turn Across Path From Opposite Directions at Signalized Junctions |
| 8 | Road Edge Departure While Backing Up | 28 | Vehicle Turning Right at Signalized Junctions |
| 9 | Animal Crash With Prior Vehicle Maneuver | 29 | Left Turn Across Path From Opposite Directions at Non-Signalized Junctions |
| 10 | Animal Crash Without Prior Vehicle Maneuver | 30 | Straight Crossing Paths at Non-Signalized Junctions |
| 11 | Pedestrian Crash With Prior Vehicle Maneuver | 31 | Vehicle(s) Turning at Non-Signalized Junctions |
| 12 | Pedestrian Crash Without Prior Vehicle Maneuver | 32 | Evasive Action With Prior Vehicle Maneuver |
| 13 | Pedalcyclist Crash With Prior Vehicle Maneuver | 33 | Evasive Action Without Prior Vehicle Maneuver |
| 14 | Pedalcyclist Crash Without Prior Vehicle Maneuver | 34 | Non-Collision Incident |
| 15 | Backing Up Into Another Vehicle | 35 | Object Crash With Prior Vehicle Maneuver |
| 16 | Vehicle(s) Turning – Same Direction | 36 | Object Crash Without Prior Vehicle Maneuver |
| 17 | Vehicle(s) Parking – Same Direction | 37 | Other |
| 18 | Vehicle(s) Changing Lanes – Same Direction | | |

Figure 10: Pre-Crash Scenario Typology covering 99.4 percent of all light-vehicle crashes for 5,942,000 cases, DOT HS 810 767, April 2017

This fact is illustrated by the pre-crash failure typology shown in Figure 10. For example, "Vehicle failure" needs further detailed and complex analysis to identify recovery policies, depending on the conditions under which this event occurs.

For autonomous systems, a key idea is to replace the individual DIR mechanisms developed at design time, by adaptive mechanisms managing system resources globally to achieve, first of all critical goals and plan best-effort goals according to resource availability. Such an approach would avoid overprovisioning of traditional approaches and would close the existing gap between critical and best-effort systems engineering [6].

Moving from correctness at design time to autonomic correctness requires not only cutting-edge theory but also finding adequate tradeoffs between quality of control and performance. The adaptive DIR process involves complex decision methods that may affect the ability to react promptly for timely recovery.

To conclude, the proposed computational model for autonomous systems can provide a basis for studying model-based autonomous system design. Nonetheless, we are far from ensuring that the conditions are in place to develop rigorous design flows.

## 5. Discussion

The main characteristic of autonomous systems is their ability to handle knowledge about their situation and adaptively respond to environment changes. The identified aspects of autonomy have some similarity with types of awareness exhibited by human mind [7].

Closing the gap between artificial and human autonomy encounters several difficult to overcome barriers.

A first barrier is that human mind understands goals in terms of high-level concepts. It is not trivial to link concepts to massive information collected by sensors or to commands of actuators. The Perception process should be robust and reliable for dynamically changing environment conditions. Similarly, there is a big distance between directives such as "deviate from the reference trajectory to avoid the obstacle" and their implementation in terms of concrete goals from which corresponding plans are effectively computed [2].

A second barrier is that situation understanding by humans is largely rooted in common sense reasoning. Our mind has built and continuously maintains since our birth, a complex semantic model of both our external and internal environments. It is practically impossible to elicit all the knowledge encompassed by such a model. No need to understand Newton's laws to expect that apple fall out of trees, that parents are older than their children are, etc. The important question is how close computers can get to a solution of this problem.
As humans have innate knowledge, we can equip an agent at design time with built-in knowledge and a faithful model of its initial environment. Then the agent's Reflection function should: 1) have access to a huge Knowledge Repository involving all common concepts and their relations; and 2) be able to consistently update the environment model by matching the perceived information to predefined knowledge patterns.

A third barrier for computers is matching human self-adaptation and the capacity: to supervise the state of acquired knowledge; to understand never encountered situations; and to create new goals. Goal creation and handling is a grand challenge of autonomy. How to assign individual goals to agents so that they all together concur to the achievement of given global system goals?

The paper provides a technical characterization of autonomy as the combination of five basic and independent features. It clearly separates aspects that are essential for autonomic behavior from other general systems engineering aspects. In that respect, it differs from other approaches using a large number of poorly understood "self"-prefixed terms: Self-configuration, Self-healing, Self-optimization, Self-protection, Self-regulation, Self-learning, Self-awareness, Self-organization, Self-creation, Self-management, Self-description [16,17]. Such characterizations based on technically non-substantiated terms obscure the debate about the very nature of autonomy.

A main conclusion is that autonomy should be associated with functionality and not with specific techniques. Machine learning is essential for removing ambiguity from complex stimuli and coping with uncertainty of unpredictable environments. Nonetheless, it can be used to meet only a small portion of the needs implied by autonomous system design. Furthermore, it is not the only way to build perceptors and controllers.

Autonomy is a kind of broad intelligence. The current AI vision is too much influenced by Turing's test that considers intelligence as a verbal game between a human and a computer. Nonetheless, animals are not verbal and exhibit intelligence. A big deal of human intelligence is not verbal.
Intelligence is not just automation of decisions even if this requires the computation of strategies with exploding complexity. Our characterization as the combination of five different types of abilities shows a big difference between an autonomous vehicle and a game playing robot. The situation awareness required for the robot is minimal. The stimuli and the environment models are trivial to interpret and build. The rules of the game are well-understood and can be directly related to goals.

Computers would exhibit intelligence when they can handle knowledge (create and use knowledge) so as to cope with the ever changing reality as humans do. Building trustworthy and optimal autonomous systems goes for far beyond the current AI challenge.

systems, Computer, January 2005.